\documentclass[acus]{JAC2003}

\usepackage[T1]{fontenc}
\usepackage{graphicx}
\usepackage{booktabs}
\usepackage{listings}
\usepackage{color}
\usepackage[scaled=.85]{beramono}

\definecolor{dkgreen}{rgb}{0,0.6,0}
\definecolor{gray}{rgb}{0.5,0.5,0.5}
\definecolor{mauve}{rgb}{0.58,0,0.82}

\begin{document}
\title{A PRACTICAL APPROACH TO ONTOLOGY-ENABLED CONTROL SYSTEMS FOR ASTRONOMICAL INSTRUMENTATION}

\author{W. Pessemier\thanks{wim.pessemier@ster.kuleuven.be}, G. Raskin, H. Van Winckel, Institute of Astronomy, KU Leuven, Belgium\\
G. Deconinck, P. Saey, ESAT-ELECTA, KU Leuven, Belgium}

\maketitle

\begin{abstract}
Even though modern service-oriented and data-oriented architectures promise to deliver loosely coupled control systems, they are inherently brittle as they commonly depend on a priori agreed interfaces and data models. At the same time, the Semantic Web and a whole set of accompanying standards and tools are emerging, advocating ontologies as the basis for knowledge exchange. In this paper we aim to identify a number of key ideas from the myriad of knowledge-based practices that can readily be implemented by control systems today. We demonstrate with a practical example (a three-channel imager for the Mercator Telescope) how ontologies developed in the Web Ontology Language
(OWL) can serve as a meta-model for our instrument, covering as many engineering aspects of the project as needed. We show how a concrete system model can be built on top of this meta-model via a set of Domain Specific Languages (DSLs), supporting both formal verification and the generation of software and documentation artifacts. Finally we reason how the available semantics can be exposed at run-time by adding a ``semantic layer" that can be browsed, queried, monitored etc. by any OPC UA-enabled client.
\end{abstract}

\section{INTRODUCTION}

\begin{figure}[htb!]
  \centering
  \includegraphics[width=80mm]{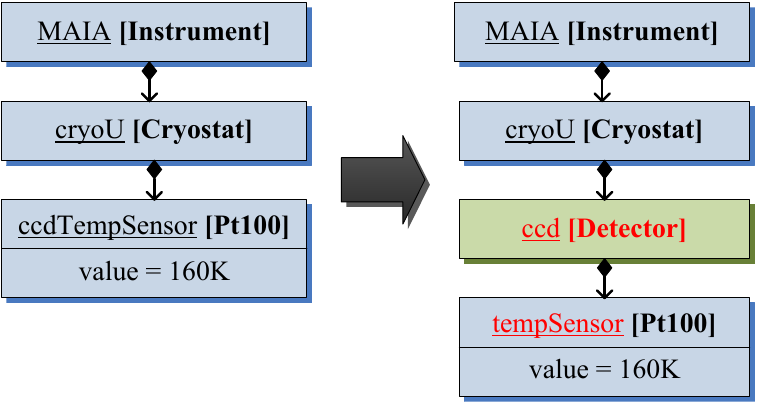}
  \caption{Example of how object-oriented models may change independently of the systems they represent.}
  \label{fig:ooproblem}
\end{figure}

Modern distributed systems to control astronomical instrumentation are typically designed as Service Oriented Architectures (SOA). Whether implemented as ``tightly coupled'' systems based on object-oriented design, or more ``loosely coupled'' systems based on a common data model, the SOA paradigm requires all parties of the distributed system to interact via shared contracts. A major drawback of these contracts is that they are often not semantically ``rich'', i.e. they are based on syntax rather than semantics. Similar to object-oriented software languages (which can only natively convey semantics such as \textit{hasType}, \textit{hasSuperClass}, \textit{hasAttribute}, ...) they lack the expressive power to model the exchangeable information in more detail. Consider the example as shown in Figure~\ref{fig:ooproblem}, which is based on the MAIA instrument. MAIA is a three-channel astronomical imager built for the Mercator Telescope and commissioned in 2013 \cite{maia}. Its control system is based on an
object-oriented design, which we are trying to improve with the results of the work described in this paper. On the left side of the figure, the temperature of the U-band detector of the instrument could be accessed over the network via the pseudocode \texttt{READ(MAIA.cryoU.ccdTempSensor.value)}. On the right side of the figure, two changes have been applied. Firstly, the detector itself (a CCD or Charge Coupled Device) is now also included in the model, and secondly the name of the temperature sensor has been changed accordingly. As a result, the pseudocode required to read the temperature is now \texttt{READ(MAIA.cryoU.ccd.tempSensor.value)}. Even though the system itself has not changed, the model has, because the model can only express structural properties using \textit{hasAttribute} relationships, and the meaning of the elements as attribute names.

\section{METHODS}

A more robust way to model the exchangeable information in a control system would be to express this information as a set of ontologies. An ontology formally represents the knowledge of a particular domain as a set of concepts, and relationships between pairs of those concepts. For instance, an ontology on the domain of \textit{electronics} (abbreviated with the prefix \textit{elec}) would define:
\begin{itemize}
    \item a vocabulary that specifies: \\
     \hspace*{2pt}--\hspace*{2pt} classes (such as \textit{Sensor} and \textit{Pt100});\\
     \hspace*{2pt}--\hspace*{2pt} instances (such as \textit{THREE\_PHASE\_POWER});\\
     \hspace*{2pt}--\hspace*{2pt} properties (e.g. \textit{senses}, \textit{powers}, ...);
    \item facts, as relationships between pairs of vocabulary terms such as: \\
     \hspace*{2pt}--\hspace*{2pt} \textit{Pt100} is a subclass of \textit{Sensor}; \\
     \hspace*{2pt}--\hspace*{2pt} \textit{senses} has \textit{Sensor} as its domain; \\
     \hspace*{2pt}--\hspace*{2pt} \textit{THREE\_PHASE\_POWER} is an instance of \textit{Power}; \\
     \hspace*{2pt}--\hspace*{2pt} Any \textit{Sensor senses} at least one \textit{Thing}.

\end{itemize}

\begin{figure}[htb!]
  \centering
  \includegraphics[width=80mm]{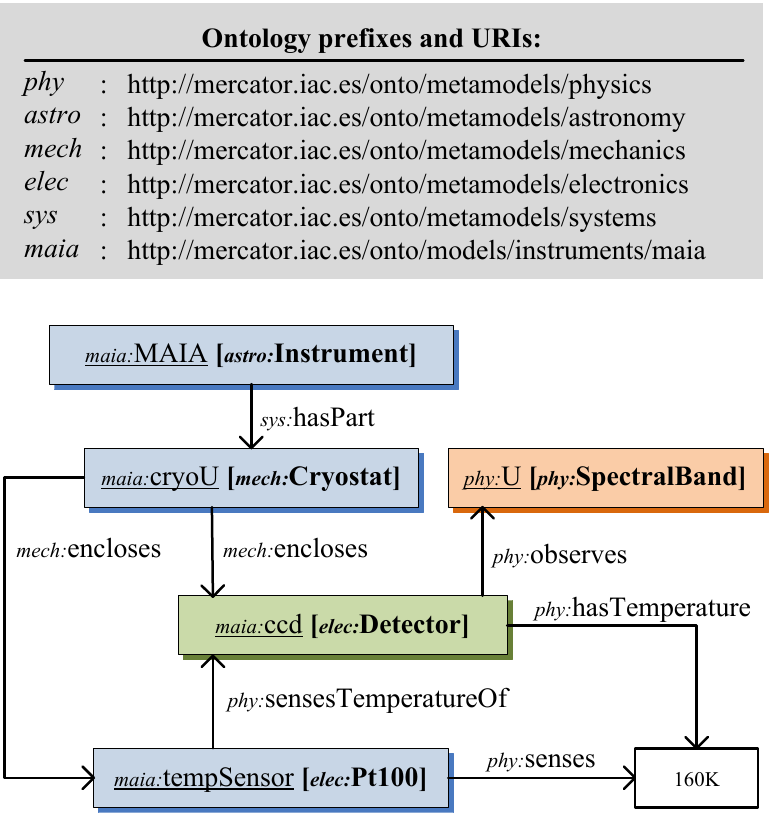}
  \caption{Same example as in Figure~\ref{fig:ooproblem}, but the system is now modeled using ontologies.}
  \label{fig:alternative}
\end{figure}

When applying this to the example from Figure~\ref{fig:ooproblem}, we can model our system using a set of ontologies, as shown in Figure~\ref{fig:alternative}.

It is clear that in our practical example, the \textit{maia} ontology only defines instances, while the properties (relations) of the instances are defined by more general ``engineering'' ontologies such as those for electronics and systems engineering. The information exposed by the model can thus be extracted by relying on the engineering ontologies (which describe generally known facts), and not on project specific ontologies. In essence, \textbf{the engineering ontologies provide a context to the application models}. When this context is shared by the applications of the distributed systen (which could be anything ranging from a historical data server to a real-time control application), the integration of applications and information sources becomes much easier and much more time-proof.

We choose semantic web \cite{semantic-web} standards to construct our ontologies, because these standards offer us sufficient expressive power to satisfy our modeling requirements, and because they are mature and come with a large set of third-party software tools. The core of semantic web technologies is the concept of \textit{subject - predicate - object} triples, which describe knowledge as facts. Subjects, predicates and objects can be identifiable via Uniform Resource Identifiers (URIs). This simple data model is defined by the Resource Description Framework (RDF), and is extended by RDF Schema (RDF-S) to offer the basic elements to build ontologies containing classes, type relationships, properties, etc. The Web Ontology Language (OWL) finally adds more expressive power to the schema, and provides the ability to specify property restrictions, cardinality constraints, inverse properties, transitive properties, etc. In the case of our MAIA example, we can use RDF, RDF-S and OWL to specify the assertions 
of Listing~\ref{lis:alternative}.

\begin{lstlisting}[caption=More knowledge captured by the ontologies.,label=lis:alternative]
sys:hasPart rdf:type owl:TransitiveProperty
sys:hasPart owl:inverseOf sys:partOf
mech:encloses rdfs:subPropertyOf sys:hasPart
phy:hasTemperature rdfs:range phy:Temperature
phy:senses owl:inverseOf phy:isSensedBy
\end{lstlisting}

\begin{figure}[htb!]
  \centering
  \includegraphics[width=80mm]{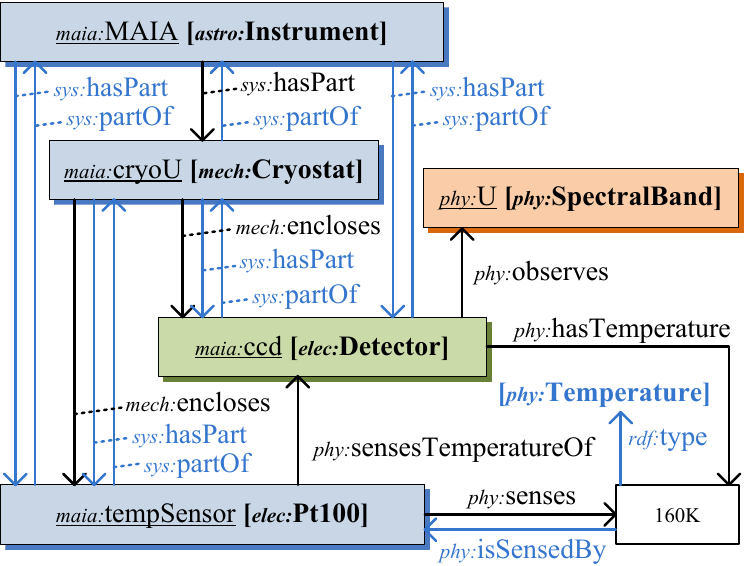}
  \caption{Implicit knowledge (blue) inferred from Listing~\ref{lis:alternative}.}
  \label{fig:implicit}
\end{figure}

A ``semantic reasoner'' can use this knowledge to draw more conclusions about our system, as shown in Figure~\ref{fig:implicit}.
Effectively, it can be inferred that the temperature sensor is now also a part of MAIA, and that, for instance, we have the freedom to use \texttt{phy:isSensedBy} instead of \texttt{phy:senses} to relate a sensor with a sensed value.

\section{PROPOSED SOLUTION}

As stated in the previous section, in our practical use case there is a clear distinction between the \textit{maia} ontology and the other, more general ontologies. We therefore discern two layers:

\begin{itemize}
 \item the ``meta-model'' layer: ontologies that contain the knowledge about the engineering domains that are involved when designing and operating astronomical instrumentation;
 \item the ``model'' layer: ontologies that contain the knowledge about the concrete applications.
\end{itemize}

\begin{figure}[htb!]
  \centering
  \includegraphics[width=80mm]{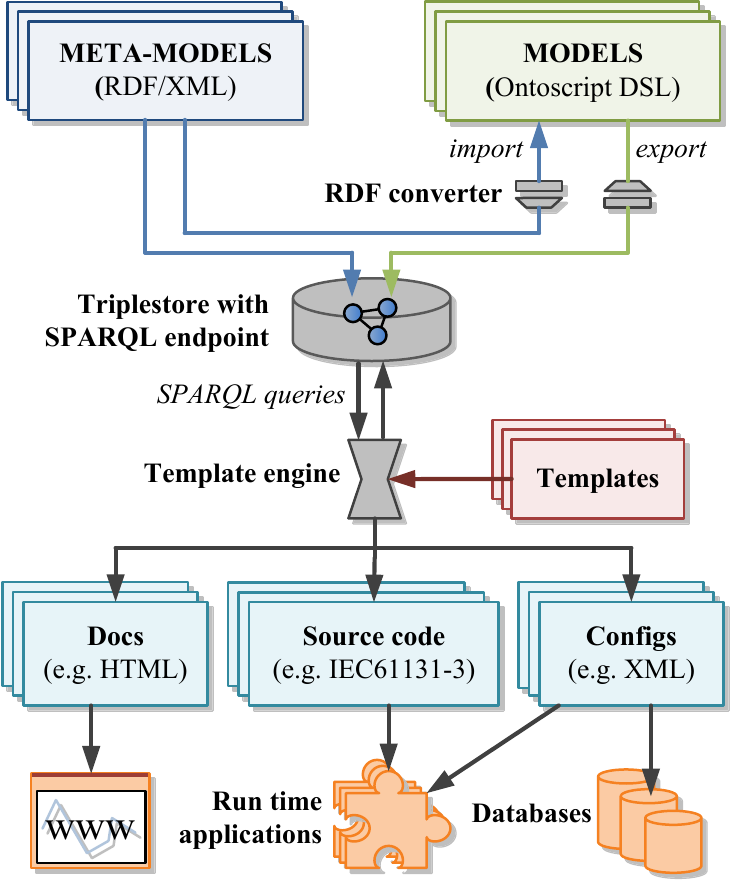}
  \vspace*{-.5\baselineskip}
  \caption{The proposed technical architecture.}
  \label{fig:architecture}
  \vspace*{-.5\baselineskip}
\end{figure}

This distinction also is apparent in the technical architecture, as shown in Figure~\ref{fig:architecture}. The next subsections will cover the different parts of this diagram.

\subsection{The Meta-model Layer}

In our approach, the meta-model layer contains the ontologies that provide the classes and their relations necessary to support the model layer. Besides a few project-independent instances (such as the \textit{U} spectral band in our example), they primarily define classes and their relations. When OWL (or rather OWL 2, the latest version of OWL) is not sufficiently expressive to model a fact, we may additionally add rules to the ontology. For instance, we could add a rule stating that ``if a \textit{Voltage} is below 120VDC, then it is an \textit{ExtraLowVoltage}''. SWRL, the Semantic Web Rule Language, would be a suitable choice to implement this rule since it is supported by some off-the-shelf reasoners. A rule language such as SWRL comes with a number of restrictions though, most noticeably that all variables in the rule should bind only to explicitly named individuals in the ontology. Since our model layer consists only of these individuals (see next subsection), this is generally no objection.

The meta-model ontologies can be considered ``heavy-weight ontologies'' because they define concepts and axioms that allow for complex inferences. As a result, we found the use of an ontology editor (preferably with integrated reasoner support) indispensable to design the meta-model layer. In our approach, we use the open source tool Prot\'eg\'e to design the ontologies, and serialize them as RDF in the eXtensible Markup Language (RDF/XML).

\subsection{The Model Layer}

In contrast to the meta-model layer ontologies, the model layer ontologies do not define additional classes or properties. Instead, they only \textit{use} the classes and properties defined by the meta-models to create new instances, and new facts about these instances. For example, whereas the \textit{electronics} meta-model ontology defines the \textit{Detector} class, the \textit{maia} model ontology defines three concrete instances of this class (i.e. one CCD42-C0 device for each cryostat).

The model layer ontologies are less heavy-weight than those of the meta-model layer, and they are developed by domain experts during the design of a particular application. Therefore we found it much more convenient to derive a set of Domain Specific Languages (DSLs) from the meta-model layer, and use these languages as a tool for building new applications. As other control system projects have demonstrated, DSLs can help to involve domain experts to build models and systems efficiently and consistently \cite{gmt-dsls}. For this purpose we have developed an internal DSL called \textit{Ontoscript} based on Coffeescript (a language that ``trans\-piles'' to Javascript). Ontoscript imports the classes, properties and individuals from meta-model ontologies so that they can be used to create new individuals and assertions about these individuals. To efficiently capture variability at the model layer level, the DSL supports parametric ``macros'' that are capable of producing similar individuals.

\subsection{The Knowledge Base}

When the meta-model and model ontologies are loaded into a so called ``triplestore'' (a purpose-built database for the storage and retrieval of RDF triples), they form the knowledge base. Off-the-shelf implementations are available that provide a database with integrated reasoner support and SPARQL-query endpoint. SPARQL is a recursive acronym for SPARQL Protocol and RDF Query Language, and is the de-facto query language for the semantic web. The integrated reasoner will make sure that, even though only explicit knowledge is imported into the triplestore (such as the model of Figure~\ref{fig:alternative}), any SPARQL client can query both the explicit and implicit knowledge (such as the model of Figure~\ref{fig:implicit}).

\subsection{The Template System}

One of the use cases of semantic web technology is the presentation of information stored in RDF triplestores via web browsers. Several frameworks exists that can query SPARQL endpoints and embed the results into dynamically generated web pages.  Template engines facilitate this task since they separate the presentation logic from the underlying business logic. They are especially useful for our purpose since not only they are capable of generating web pages, but also source code files, configuration files, CSV (Comma-Separated Values) files that can be imported by spreadsheet software, etc. We found versatile template engines such as Mako very suitable since they allow us to embed SPARQL queries within the templates.

An interesting application of semantic modeling in the context of system development is the formal verification of the system design. As others have shown, system verification is often the result of fairly simple reasoning \cite{verification}. OWL, and especially SWRL, are in many cases expressive enough to model design requirements. In case of MAIA, we could define a meta-model called \textit{maia-requirements} stating that all detector temperatures must be measured by a sensor and exposed by a communication middleware. A simple SPARQL query can verify this requirement, and can be embedded in an HTML (HyperText Markup Language) template to produce a web-based report.

\subsection{The Run Time System}

In order to benefit from the proposed ontology-enabled architecture, the software applications must be able to interact using the same context, as provided by the meta-model ontologies. We identify two implementation scenarios:

\begin{itemize}
 \item the context information is only available at ``compilation time'' (i.e. when the software artifacts are generated from the knowledge base via the template system);
 \item the context information is also available at ``run time''.
\end{itemize}

In the first case, a change in the model layer will result in the generation of new source code artifacts. Other source code or compiled software that depend in some way on these artifacts may be affected as a result. The benefit compared to a traditional model-based approach is however that the templates that generated the artifacts do \textit{not} have to change, as they can construct their SPARQL queries primarily using meta-model terms. For example, we could define a template that generates Python code to retrieve the temperature value of the MAIA U-band detector via OPC UA communication (Object Linking and Embedding for Process Control, version Unified Architecture \cite{opcua}). Based on the model of Figure~\ref{fig:implicit}, an excerpt from the template is shown in Listing~\ref{lis:mako}. In a more realistic scenario we would define helper functions to extract frequently required information such as communication middleware details. Although any middleware can be modeled in our ontologies, OPC UA has 
the advantage that it is capable of representing RDF-like triples. Similar to semantic web technology, OPC UA can express complex graphs with fully qualified nodes and binary relationships.

\begin{lstlisting}[caption=Mako template to retrieve the temperature value(s) of the U-band detector over OPC UA (by using our open-source Unified Architecture Framework or ``UAF'').,label=lis:mako]
<% results = sparql.simpleQuery("""
SELECT ?svrUri ?nsIdx ?id WHERE {
  ?det    phy:observes            phy:U       .
  ?det    phy:hasTemperature      ?temp       .
  ?temp   opcua:hasExpandedNodeId ?nodeId     .
  ?nodeId opcua:hasServerUri      ?svrUri     .
  ?nodeId opcua:hasNamespaceIndex ?nsIdx      .
  ?nodeId opcua:hasIdentifier     ?id  } """) %>
def getUTemperatures():
  addresses = []
  % for r in results:
  addresses.append(Address(
    NodeId(${r.nsIdx},"${r.id}"),"${r.svrUri}"))
  % endfor
  return UAF_client.read(addresses)
\end{lstlisting}

The second case goes one step further in the sense that the semantic information from the meta-model layer is now being used at run time. For instance, the databases that store and serve historical sensor data could be RDF triplestores with SPARQL endpoints instead of relational databases. However, when entering the application domain of real-time control and data acquisition, we notice that the use of semantic web technology quickly reaches its limits. The HTTP (HyperText Transfer Protocol) communication on which it is based is generally not sufficiently efficient or deterministic for control applications. In contrast, the industrial OPC UA standard does provide an efficient binary protocol, and is also capable of exposing the semantics of our meta-model and model ontologies. OPC UA additionally provides services to query, browse, read, write, modify, etc. the exposed graphs, and it is readily available on industrial platforms such as Programmable Logic Controllers (PLCs).

\section{CONCLUSIONS}

Having experimented with a prototype set-up of the architecture as shown in Figure~\ref{fig:architecture} and as applied to the MAIA instrument, we found a significant added value of applying ontologies to the design of distributed control systems for astronomical instrumentation. By strictly separating the ontologies that provide context from the ontologies that describe concrete applications, we expect that the produced control systems will be more evolvable. We think that, for practical reasons, this separation of concerns should be extended to the tools used to engineer the ontologies: dedicated ontology editors may be most suited to model the ``meta-model'' layer, while Domain Specific Languages may be more suited for describing the systems itself. Because the focus of our research is to semantically integrate the different applications and data sources in a distributed control system, the ontologies should primarily support the organization and exchange of information. Semantic web technologies such as
RDF, RDF-S, OWL, SWRL and SPARQL are specifically designed for this purpose in mind, albeit in a different field of application. When semantic technology is required at the lower levels of the control system architecture, we argue that OPC UA may be a natural choice to expose the knowledge even at run time.

\end{document}